# "Spreading Groups" in Twitter: The Flow of Rumors about Stocks

ALON SELA1, Tel-Aviv University
ORIT MILO-COHEN, Ben-Gurion University
IRAD BEN-GAL, Tel-Aviv University
EUGENE KAGAN, Ariel University

The paper addresses a method for spreading messages in social networks through an initial acceleration by "*spreading groups*". These groups start the spread which eventually reaches a larger portion of the network. The use of spreading groups creates a final flow which resembles the spread through the nodes with the highest level of influence (opinion leaders). While harnessing opinion leaders to spread messages is generally costly, the formation of spreading groups is merely a technical issue. The paper presents an information flow model and inspects the spread in a dataset of Nasdaq-related tweets.

・Information systems → World Wide Web → Web applications → Social networks ・Information systems → World Wide Web → Online advertising → Social advertising ・Information systems → Information systems applications → Collaborative and social computing systems and tools → Social networking sites

## 1. INTRODUCTION

Modern social network platforms provide a simple and efficient way for spreading messages among individuals or communities of users. The results of such messages spread (the terms 'messages' and 'information' are used interchangeably) can vary in scale: from serving the local need of an individual user up to a larger influence on political, social and economic phenomena, such as the "Arab Spring" (Howard, et al., 2011) or the "Occupy Wall Street" (Bennett, et al., 2011) movement. In fact, in several areas the role of social networks and the activities of small groups of users, can lead to an outcome, which can be as influential as that of a traditional media.

One of these areas is the financial market, where news' spreading through social network are, in many cases, published before these news reach the "traditional" media (Leskovec, et al., 2009). The stock markets, are fundamentally of a dynamical nature, and allow quantitative analysis of such news while using direct objective measures such as gains.

Studies of internet messaging influences on financial markets can be tracked back to the rise internet *message boards* and *forums*. In early 1998, for example, the correlation between the number of published messages on companies in Yahoo Finance Board and their shares' values was noticed by the US Securities and Exchange Commission (Commission, 1998). The social influence in the stock markets was further studied by (DeMarzo, et al., 2003) to suggest a model of a network activity and its related influence. The authors argued that the users form their opinions with respect to opinions of other users, while being biased by the appearance of repeating messages, and to a less depending on the accuracy of the information presented in these messages. The value of information in financial

Author's addresses: Alon Sela, Department of Industrial Engineering, Tel Aviv University; Orit Milo-Cohen, Guilford Glazer Faculty of Business and Management, Ben-Gurion University; Irad Ben-Gal, Department of Industrial Engineering, Tel Aviv University; Eugene Kagan, Department of Industrial Engineering, Ariel University.

markets was further studied by (Cao, et al., 2002), who claimed that financial traders might be less affected by the actual stock costs, rather than by the intensive conversations about the stock. Following this direction, it was demonstrated that a higher message posting might predict negative stock returns, higher volume and higher volatility (Antweiler, et al., 2004). This opinion coincides with previous findings (Harris, et al., 1993) who argued that opposite opinions among posted messages tend to be followed by increased trading volume. Thus, an increase in the message spread by itself, regardless of its content, might be of a financial value.

With the development of communication technology, the message boards were mostly replaced by instant messaging platforms and social networks, such as Twitter and WhatsApp that, in addition to the ability to accelerate and facilitate the communication between the users, introduced a new grouping feature which enables the creation of messaging groups and network structures.

In this paper, we present a model of information flow in Twitter and analyze spreading of news and rumors related to stocks, which contain the NASDAQ-100 hash tags. More precisely, we analyze the potential influence that can be achieved by an organized group that intentionally spread messages, and present some empirical results that support the actual existence of such spreading groups.

The rest of the paper is organized as follows. Section 2, introduces a model of message-spreading through network groups, defines some of its characteristics and formulates criteria for the analysis of such an information flow. Section 3 describes the raw data obtained from Twitter messages and outlines the analysis method. Section 4 presents the obtained results from both simulated studies as well as real data support. Section 5 discuses some of the consequences of the observed phenomena, including recent implications on bots of message spread. Section 6 concludes the paper.

## 2. THE MODEL OF THE NETWORK AND INFORMATION FLOW

The proposed model follows the general approach of information flow in social networks, as shown for example in (Kempe, et al., 2003) and (Newman, 2010). It implements the statistical properties of such a flow over Power Law graphs (Bollobas, 2001). The model implements a retention loss dynamics (e.g., see, (Weng, et al., 2012) and (Hirshleifer, et al., 2003)), by which the probability of a retweet decreases over time, while measuring the final number of nodes which received the message. The model compares the value gained from the creation of an intentional spreading group to different policies for initial information spread, and provides insights for further understanding the results obtained from real Twitter messages on the NASDAQ-100 stock market.

### 2.1 Network Structures

Let $G = (V, E)$ be a graph that represents the Twitter (or similar) social network. The set of vertices $V = \{v_1, v_2, \ldots, v_n\}$ is associated with the individual users in the network, and the set of edges $E = \{e_{ij} = (v_i, v_j) \mid v_i, v_j \in V; i, j = 1,2, \ldots, n\}$ defines the connectivity between the users.

In general, in the models of networks it is assumed that $G$ is finite, but the number $n$ of vertices or the number of edges (or both) can vary over time. For our purposes, consider the case with constant $n$ and varying number of edges. Following the *preferential attachment* model (Barabasi, et al., 1999), let us assume that new nodes are added to an initially sparse graph $G$. The probability that a new vertex $u \in$ $V$ will be connected to vertex $v_i$ depends on the connectivity $k_i$ or the degree of $v_i$, representing the number of vertices with which $v_i$ is connected, i.e.,

$$p(v_i) = \frac{k_i}{\sum_{j=1}^{n} k_j}, \quad i = 1, 2, \ldots, n.$$

A *Spreading Group* of users is denoted by a sub-graph $G'$ of the graph $G$ that is defined as follows. Let $G' = (V', E')$, $V' \subset V$ be a subset of vertices of size $\#V' = \lceil nr \rceil$, where $r \in (0,1)$ is a given ratio value and $\lceil a \rceil$ stands for the ceil of the number $a$. During the construction of $G$, edges between pairs of vertices from $V'$ are added arbitrary in such a manner that the average connectivity of the sub-graph $G'$ increases *independently* of the connectivity of $G$. As a result, the graph $G$ includes a sub-graph $G'$ with a higher connectivity; such a structure corresponds to the network with distinguishable subset of interconnected users.

### 2.2 Message seeding and information flow

As indicated above, the construction of graph $G$ follows the preferential attachment model and a distinguishable Spreading Group represented by the sub-graph $G'$. Now let us consider the initial message seeding and information flow in the network. Denote by $V_s \subset V$ a subset of vertices, which are associated with network users that initiate the messages seeding. For the purpose of the presented analysis, consider the following three types of initial spread options for selecting the seeding vertices:

(1) The seeded vertices $v \in V_s$ are selected randomly from the set $V$ independently whether they are in the spreading group $V'$ or not.
(2) The seeding vertices $v \in V_s$ are selected randomly from the set $V'$ of vertices, which is associated with the Spreading Group.
(3) The set $V_s$ of seeding vertices is specified as a set of vertices with the highest Eigenvector Centrality measures (Newman, 2010); in the other words, $V_s$ is associated with the group of users with the highest influence in the network.

After the initial seeding by vertices $v \in V_s$, information is spread according to the *Independent Cascade* model (Kempe, et al., 2003). The model assumes that messages can be transmitted from a user to its neighbors. The receiving user can then spread the message to its neighbors with a transmission probability $p(t)$. Unlike the independent cascade model, where a node can transfer a message only in the consecutive period of time, we followed the analysis of real Twitter messages in our data set, as well as the theory of the limited attention of the users (Weng, et al., 2012), and assumed that the transmission probability decreases in time. The initial transmission probability for each node during the first time unit after receiving the message is $p_0 \equiv p(t = 0)$, and is recursively set in the consequent time to be

$$p(t + 1) = \frac{p(t)}{c}, \quad t = 0, 1, 2, 3 \ldots,$$

where the retention loss factor in the simulation runs took a value within the range $c \in (1, 10)$. In other words, when a user receives a message, the probability of spreading it in the first time period is set to a maximal value $p_0$, which then decreases over time causing the process of information spreading to continue until the network reaches a final state, as was defined by a state where the number of users who receive the information does not change in two consecutive time steps.

The above model was analyzed by a simulation study implemented in the NetLogo simulation platform (Wilensky, 1999). The simulations were conducted using a network of $n = 10,000$ vertices. In these simulations, different seeding methods were compared; i.e. (1) seeded vertices $v \in V_s$ while nodes selected randomly from $G$, (2) seeded vertices $v \in V_s$ while nodes selected randomly from $G`$, (3) seed nodes with starting from the node with the highest Eigenvalue Centrality measure, then to the 2$^{nd}$, 3$^{ed}$…high Eigenvalue centrality. The model`s parameter were modified such that the dense cluster $G`$ contained 1%, 3% or 5% of nodes in $G$. Second, the number of initial seeds, $|V_s|$ was set to 15, 25 or 35 nodes, respectively. Third, the transmission probability $p_0$ was set to 1%, 5% or 10%. Last, the retention loss factor c was set to 1.5, 3 or 4.5.

Given the above-mentioned set of parameters, the simulation study included 4,860 runs (20 runs with each distinct combination of parameters), while in each trial the total number of the vertices that were exposed to the message by the end of the trial was recorded and analyzed. The average numbers of exposed vertices over the trials for the different sets of seeding methods and number of initial seeds are shown in Figure 1.

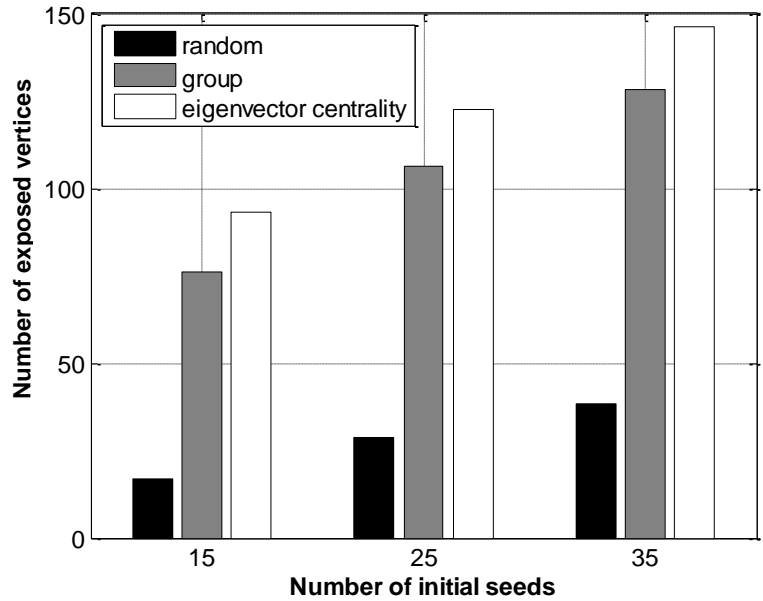

Fig.1. Average number of exposed vertices for three types of seeding methods and for number of initial seeds. Black bars correspond to the seeding vertices selected randomly from the set of all vertices; gray bars correspond to the seeding vertices selected from the spreading group and white bars correspond to the seeding of vertices specified by the vertices with the highest Eigenvector Centrality measure.

Note that if the seeding vertices are selected randomly from the set of vertices, then the number of the exposed vertices at the end of the simulation is nearly three times smaller than this number in the case that the seeding vertices are selected from the spreading group. Such result coincides with the findings from the evaluation of the twitter messages as presented in the next section, which present possible existence of such spreading groups. Since the connectivity in the spreading group $G'$ is higher than the connectivity in the network $G$, the messages in $G'$ circulate more intensively and, consequently, are better exposed to users in the remaining network. In addition, let us note that the used information flow model, does not consider the meaning or the content of the messages (Harris, et al., 1993), nor the increased tendency to read a message arrives from different users in the social circle (herd behavior). These influences, can further stress the difference between the structures of the random seeds as compared to the ignition of the message by seeding groups as described above.

## 3. EVALUATION AND ANALYSIS OF TWITTER MESSAGES

In this section, we follow the above information-spread insights and analyze real Twitter messages on the Nasdaq-stock market. Characteristics of message circulation in the network are used for recognizing possible spreading groups' structures.

### 3.1 The data

The row data included 1,481,444 "NASDAQ-100 hashtags" tweets that circulated in the Twitter throughout June 2014. A filtering process was applied to select tweets that include the '$' sign before the identifying string. Based on Twitter jargon, such a selection distinguishes between the stock itself (e.g., the $AAPL stock) and other possible objects with the same name (e.g., the fruit or the company that carry the name 'Apple'). As a result, the obtained filtered set of tweets contained 250,937 unique tweet messages.

Each unique record in the data set represents a single occurrence of a tweet. Together with the tweet text, the record included a field tag indicating the time it was published; the name of the publisher; whether the tweet was retweeted, and, in cases of retweets, the identifier of the user who initialized the tweet cascade. This method of data presentation hides the network structure, a valuable resource with a commercial value, which is a part of the Twitter commercial assets.

At the next stage, repetitions of the same messages were analyzed to identify that among the 250,937 tweets only 5,665 repeated more than once. Thus, in our data set, approximately 2% of the tweets, which include the NASDAQ-100 stock hash tags during June 2014 were repeated leaving 98% of these tweets unrepeated. The distribution of the tweets with respect to the number of their repetitions is shown in Figure 2.

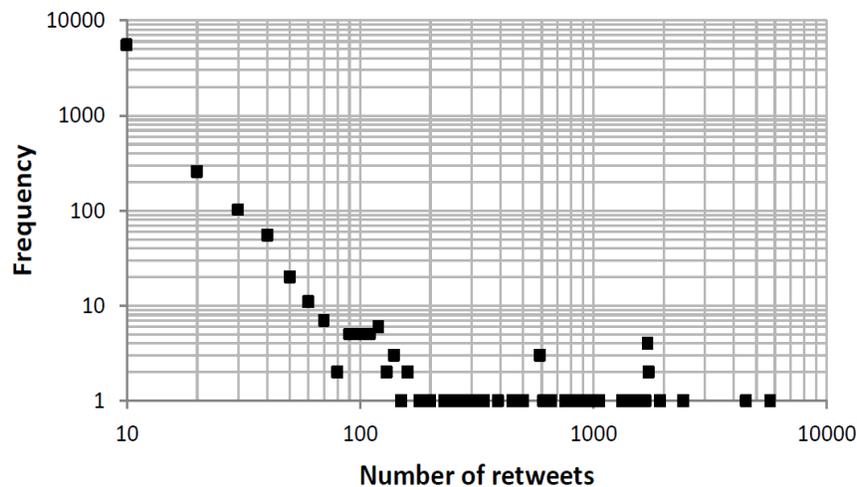

Fig. 2. Distribution of the re-tweets with respect to the number of their repetitions (on log-log scale). The distribution follows a power law up to a repetition level of approximately 300 repetitions. Further repetitions might be following a different process.

Note in Figure 2 that the retweets frequency, at the initial spread up to 250-300 repetitions, follows a power law distribution in most cases, while having few outlier tweets that demonstrated an higher-than-expected level of repetitions. Such anomaly phenomena might be explained, for example, by the repetitions of important messages that were initiated by the users' guru(s) or by the circulation of messages spread by some other methods, such as a spreading group or a bots (botnets).

### 3.2 Types of message spreaders and expected spreading scenarios

In this section we consider several possible scenarios, by which information can spread through networks. The first scenario, called 'naïve spreader', corresponds to a case in which an unknown user posts an interesting new message in Twitter and causes a large information cascade. The user has no aim or invests no effort to manipulate and increase the spread.

The second scenario, called 'guru spreader', is associated with a 'guru' being followed by his followers. In this scenario, an initial message twitted by the known user ('guru') is spread by a number of users ('followers'), who find this message interesting. Such a case might be seen for example in recommendations of gurus such as Warren Buffett.

The third scenario, on which we focus, corresponds to the 'spreading group', where a distinguishable group of users intends to spread a message by high repetitions within the group. Such a strategy might hide or present the initial spreader, but is expected to contain messages of lower economic value (since their source is not an expert or a guru).

As demonstrated in Figure 1, such a message-flow scenario results in a spread that resembles the spread through the highest influential users in the network. While these highly influential users are usually hard to reach, since they are celebrities or opinion leaders, the formation of spreading groups is no more than a technical issue but can yield similar results.

Distinguishing between the scenarios of a guru followed by a group of followers, and that of a spreading group, is not always easy. A guru might create a group of followers which form a cluster with a higher connectivity. The formation of a dense cluster around a financial guru is only a result of his financial wittiness. In order to inspect if the clusters are formed around a financial guru or are only a method to increase the messages spread, we concluded an evaluation of the message`s content, as would be further explained in section 4.2. below.

### 3.3 The Twitter data and methods of analysis

As indicated above, among the 250,937 tweets, which include the NASDAQ-100 stock hashtags, we selected 5,665 messages that were found to be repeated more than once. These highly-repeated messages were separated into three subsets: a subset of tweets with more than 700 repetitions, a subset of tweets that repeated between 100 to 400 times, and a subset with tweets repeating between 401-699 times.

The subset of highly-repeated messages contained 79 unique tweets that repeated 2,606 times per message on average. In this subset, two messages with extremely high numbers (outliers) of repetitions (57,026 and 27,57) were omitted, leaving 77 highly-repeated tweets. The group of lower number of repeated messages contained 143 unique tweets that repeated only 173 times per message on average. The middle subset, of tweets repeating between 401-699 times was ignored, thus forming two distinct groups, i.e. the highly repeating messages and the lowly repeating messages.

To obtain a lowly repeating group of the same size as the group of highly repeated messages, in each run, 66 randomly chosen tweets were omitted. As a result, the analysis was conducted over groups of similar sizes, each of which contained 77 unique tweets. A repetition of the process of randomly omitting users was performed several times, to ensure this random choice does not influence the final results.

A comparative analysis of the messaging activity for the users in both the highly and the lowly repeated messages was executed, including a comparison of the rates of reoccurrence of similar users and the distributions of the numbers of tweets in each group. In addition, the relation between the messages spread and the messages` 'financial-importance' was evaluated by 97 independent reviewers. Such an analysis enabled a better separation between spread of rumors and spread of meaningful messages, which enables the separation between a scenarios of a 'guru spreader' to the scenario of the 'spreading group'.

## 4. RESULTS

The analysis addresses the reoccurrence of the tweets and distributions of users in both groups of highly and lowly repeated messages, as well as the content of messages in these groups. The main objective of the analysis is to identify messages repetitions with respect to the users involved in the spread of these messages, and in relation to their content.

### 4.1 Rate of the tweets reoccurrence

The analysis is conducted using the groups of highly and lowly repeated tweets; as indicated above both groups contained 77 unique tweets with an average of 2,606 and 173 repetitions (retweets) respectively. For both groups, the users involved in the messages spreading were identified; the records of such users in the groups of highly and lowly repeated messages are denoted by $\vec{V}_h$ and $\vec{V}_l$, respectively. Notice that since the same messages can be sent by different users and, alternatively, different messages can be sent by the same users, each of the records $\vec{V}_h$ and $\vec{V}_l$ might include repeating user identifiers. The lengths of the obtained records were $\#\vec{V}_h = 204{,}442$ and $\#\vec{V}_l = 16{,}919$.

In order to have a basic outlook on the characteristics of users from the records $\vec{V}_h$ and $\vec{V}_l$ the following process was performed. (1) A random sample of 200 users was chosen from each group by sampling 10 users from a sample of 20 highly and lowly repeated tweets, respectively. (2) For each of the 200 users, all the tweets, in which this user was involved, were collected. (3) The data was filtered by removing tweets with no repetitions. (4) The repetitions of names of users in these two groups were plot on a histogram for both the highly and lowly repeating group. The distributions of the numbers of tweets repetitions for these two groups of users are shown in Figure 3, where Figure 3.a presents the users from the highly repeating tweets, and Figure 3.b presents the users from the lowly repeating tweets.

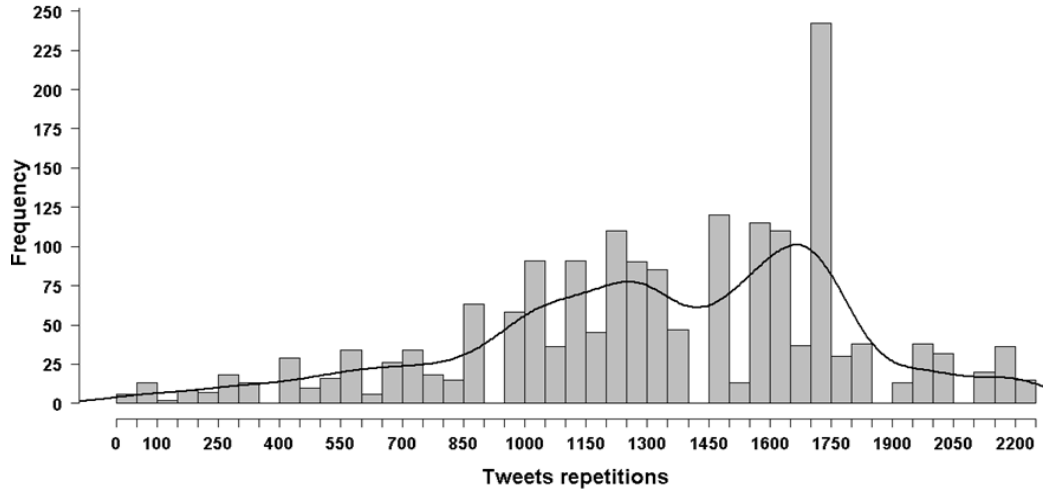

(a)

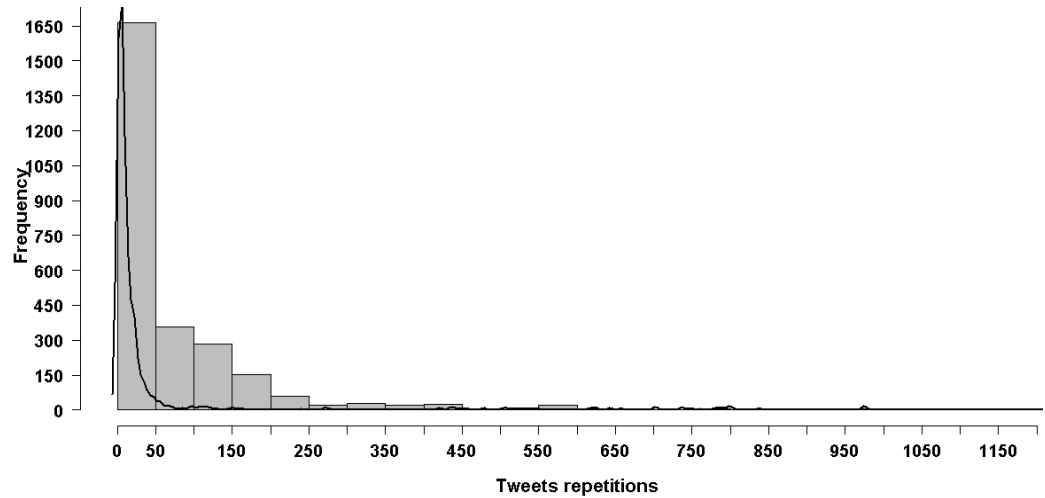

(b)

Fig. 3. Distributions of the numbers of tweets repetitions for the groups of users involved in spreading highly (a) and lowly (b) repeated messages. Figure (a) depicts the frequencies of tweets repetitions for the users that have at least one tweet that repeated more than 700 times; i.e. group of highly repeating messages. Figure (b) shows the same frequencies for the users that have at least one tweet that repeated between 100 and 400 times; i.e. group of lowly repeating messages.

Note that the retweets distributions of users from the 'highly repeating' tweets Figure 3.a) and the 'lowly-repeating' tweets (Figure 3.b) are significantly different. Moreover, the lowly-repeating' tweets distribution follows the power law as expected for the information flow that do not contain the 'spreading group' scenario (cf. Figure 2).

Note that both groups of lowly and highly repeated messages might contain tweets sent by both naïve users and organized 'spreading group', which are mostly involved in the initial stages of the spread. Thus, in order to separate the 'spreading group' from the 'naïve spreaders' scenarios, for each tweet we identified the $m$ earliest spreaders in the records $\vec{V}_h$ and $\vec{V}_l$ and compared the rates of repeating names of the spreaders. The records of such earliest spreaders for the highly and lowly repeated messages are denoted by $\vec{V}_h^m$ and $\vec{V}_l^m$, respectively. Finally, by $V_h^m$ and $V_l^m$ we denote the sets of users appearing in the records $\vec{V}_h^m$ and $\vec{V}_l^m$, correspondingly.

The rates of the tweets reoccurrence for the groups of highly and lowly repeated messages were calculated as follows:

$$R_h^m = \frac{\#\vec{V}_h^m - \#V_h^m}{\#\vec{V}_h^m} \quad \text{and} \quad R_l^m = \frac{\#\vec{V}_l^m - \#V_l^m}{\#\vec{V}_l^m}.$$

It is clear that if each user only sends a single message, then the corresponding rate is equal to zero, and if all the messages were initially sent by the same users, then the rate is close to one.

Let us now consider the dependence of the rates $R_h^m$ and $R_l^m$ on the number $m$ of the earliest spreaders, as found from the real Twitter's data. These dependences are shown in Figure 4.

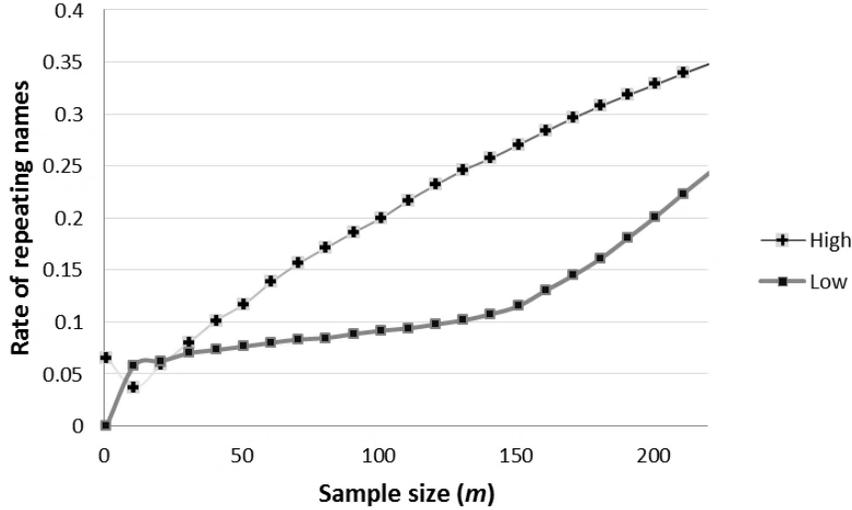

Fig. 4. Proportion of repeating names for m earlest users for highly and lowly repeated tweets.

It is seen that for a small $m$ the numbers of repeating users in both groups is also small and coincides with low probability of reoccurrence of the messages. However, as $m$ increases, the difference between the numbers of repeating users grows, and for $m$ close to 200 in the group of highly repeated messages this number is nearly twice higher than in the group of lowly repeated messages. Furthermore, one would assume that if no spreading groups exist, than the highly repeating messages group should include a lower number of repeating names (since the sample $m$ is of a larger number) compared to the lowly repeating messages group. Note, however, that the opposite case is observed in the real data. Finally, note that the rate of repeating users in the beginning of the spread of the highly repeating messages group, is higher than that rate in the lowly repeating messages group.

### 4.2 Content of messages and messages spreading

As it follows from the distribution of retweets in Figure 3.b, the information flow by the users from the record $\vec{V}_l$, who are the users involved in spreading lowly-repeated messages, is the same as the distribution of the messages spreading by the naïve users or random spreaders (see section 3.2). One could argue that the highly-spread messages are often a consequence resulting from an important and valuable message, which spread to a higher number of users due to its valuable content. Such a case might fit the scenario of a financial guru(s), such as Warren Edward Buffett, whose messages are highly important, and thus quickly spread. In order to inspect this phenomenon, an analysis of the tweets spread with respect to their content was performed. The comparison included an inspection of the spread from one typical spreader from $\vec{V}_h$ and one typical spreader from $\vec{V}_l$. The analysis enabled the evaluation of messages importance in the context of their degree of spread.

For such cases, we identified some well-known users, of high significance in the financial market that appear in the data set, such as CNBC, ForbesTech, FortuneMagazine or BloombergTV. Similarly, on the list of the most retweeted messages, we also identified some less-known users, which are not considered as influential on the considered market, such as teacuppiglets, philstockworld or 2morrowknight. In both of these groups, we examined the numbers of tweets and retweets, as shown in Table 1 (sorted by the average number of retweets).

Table 1. The numbers of tweets and retweets for well-known and unknown users

|  | User name | Total Number of Tweets | Number of retweets | Average number of retweets |
|---|---|---|---|---|
| Well-known users | CNBC | 113 | 4,781 | 42 |
|  | ForbesTech | 36 | 1,096 | 30 |
|  | CNBCnow | 25 | 736 | 29 |
|  | BloombergNews | 23 | 575 | 25 |
|  | FortuneMagazine | 57 | 1,186 | 21 |
|  | YahooFinance | 87 | 1,606 | 18 |
| Less-known users | philstockworld | 35 | 45,198 | 1,291 |
|  | Teacuppiglets | 44 | 31,240 | 710 |
|  | WSJ | 112 | 13,949 | 125 |
|  | androsForm | 32 | 3,507 | 110 |
|  | ValaAfshar | 26 | 1,174 | 45 |
|  | 2morrowknight | 35 | 1,073 | 31 |

Unlike one might expect, note that messages posted by the well-known financial firms are less retweeted compared to the messages posted by the less-known users. For example, the average numbers of retweets of the messages posted by CNBC and ForbesTech are, respectively, 42 and 30, while the average numbers of retweets of the messages posted by philstockworld and teacuppiglets are, respectively, 1,291 and 710. As another example, the total numbers of tweets and retweets and average number of retweets per message for the respectable company ForbesTech (the second line in the table) is approximately the same as these numbers for the user 2morrowknight (the last line in the table).

A high-level comparison of the content of the top messages sent by well-known firms, such as BloombergNews, ForbesTech and YahooFinance, to the content of the messages sent by less-known users demonstrates that the tweets by the former users often include fundamental information that can be relevant to investors, while the tweets by the latter users had, in many cases, a potentially lower content values. For example, the spread of the message "*RT@teacuppiglets: Bought 6000 shares of $fb here lets go:)*", was spread by 1,051 users, and might represent (with some doubts) a group of investors following a financial guru. However, the spread of the messages "*RT@teapartymobile: "$AAPL about to have a huge day! Gapping up already! Weeeeeeeeeeeeeeeeeee!!! $MLCG $FB*", was retweeted by 1,726 users, or the message "*RT @teacuppiglets: Our private $FB investors group has almost reached 1500 members. There is power in numbers! Free 2 join it is free https…*" was retweeted by 1,575 users, thus, might be considered as a spreading group artifact.

### 4.3 Evaluation of message importance with conjunction to its spread

The evaluation of the tweets' content in conjunction to their spread was performed by a survey, aiming to inspect the subjective importance of different tweets. The survey was answered by 97 Mechanical Turk workers, each evaluating a total of 10 tweets regarding their importance. The importance was in a 1-5 scale, and the set of tweets contained the 20 most spreading tweets of two chosen members of table 1. The first member was selected from a set of the well-known users, while the second member was selected from the set of unknown users. These 40 evaluated tweets were randomly divided to 4 different surveys, while the name of the spreader was hidden in the tweet in order to prevent any bias (negative or positive) towards the well-known financial firm. The final results in Figure 5 contained 970 evaluations of 40 different tweets, as performed by 97 users.

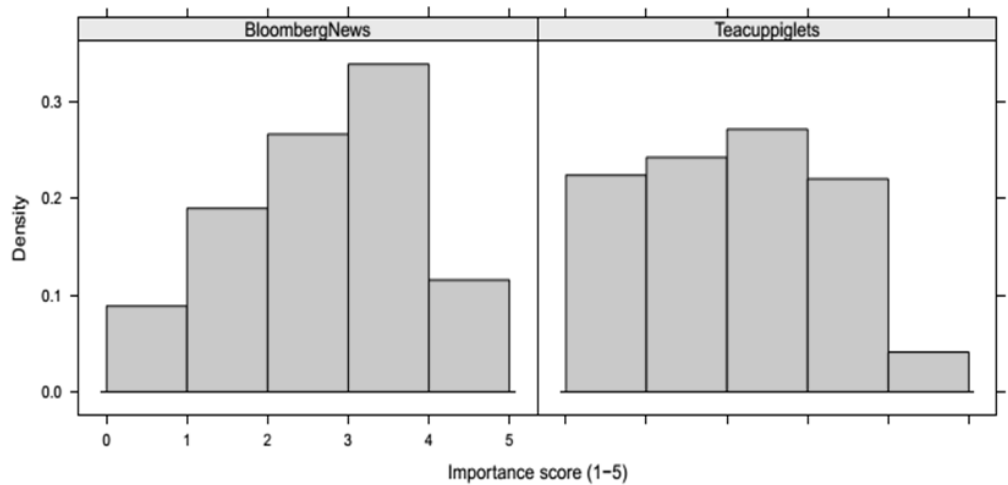

Fig.5. Messages importance evaluation of two information sources: A well-known source - BloombergNews (left); and a less-known source - teacuppiglets (right). The comparison was concluded by 970 evaluations performed by 97 distinct reviewers and reveals a potentially greater importance for the messages that were spread by the well-known source.

The obtained results further support the assumption that the levels of spread of the sources (mean retweet of 710 for teachppiglets vs. 25 for BloombergNews) are not necessarily related to the content importance of their messages.

### 5. DISCUSSION

The existence of social bots which mimic human behavior has been known for quit some times. Their effect and influence on political, social and economic aspects has reached such a degree, that DARPA, the American Defense Advanced Research Projects Agency, held a 4-weeks competition in which multiple teams from its Social Media in Strategic Communications program competed to identify a set of previously identified "influence bots" in Twitter (Subrahmanian, et al., 2016). These groups, were competing on nearly 4 million USD prize, and included highly skilled data scientists, focused on identifying bots by (1) their semantics, (2) temporal behavior, (3) anomalies in features of the users profile and (4) the user's topological deviation from known network features. While the success rate (accuracy) of correct bot/human classification for the winning team has reached as much as 97.5%, the third team only reached a 90% accuracy. Nevertheless, these encouraging results for bot/human identification do not shed the concern of these bots activities. They are only a temporary step in the competition between bots and the naïve users' activities. While the exact winning algorithm is probably kept by DARPA, some current versions of similar bots detectors can be freely found in work by Davis et. al (Davis, et al., 2016).

While DARPA`s winners seem to differentiate between social bots and humans, this is not the case for the usual naïve users nor the social network sites themselves. An interesting result, obtained in (Freitas, et al., 2015) demonstrate well the difficulties of such a classification. Freitas constructed 120 social bots and followed their social development throughout time. All these bots had a profile, followed other users, and generate tweets. They soon started to be followed by naïve users. Moreover, only 31% of the bots were detected by Twitter within a period of one month, at the time that the article was publishing. More interestingly, as much as 3 of these bots reached similar levels of popularity and influence (as measured by the "Klout score" (Anger, et al., 2011), which is a common social networks influence measure) as some top data scientists in Facebook, and even surpassing a principal research member from IBM. These results demonstrate again the growing influence of social bots, and the relevance of an evaluation measure of their activities as well as the formation of awareness to their strategies.

While detecting a single social bot, seems to be a task in hands for an organizations similar to DARPA, we believe the next level of complexity of bots evolution would include the use of networks of bots as described in this work. These bots' networks, can be 'mixed' with human users, either naïve or not, to further distort the ability of detection).

This work demonstrates the effectiveness of strategic sub networks of users that can promote information spread. It is only a matter of time until networks of bots would collaborate in order to reach a greater level of influence while their detection becoming harder.

It is only natural that with the development of AI technologies, as well as the increasing importance of social networks we will see further growth in the complexity of networks of social-human bots.

## 6. CONCLUSION

Social networks are an important aspect of modern information spread. Their influence on social, economic and political aspects of our lives is evident. While a common practice used by commercial organizations to increase their influence and popularity is focused on harnessing influencers (e.g., celebrities), the engagement of these celebrities to promote commercial products (or social agendas) is usually costly and dependent on their personal preferences.

This work presents the advantages of a more technical (while probably less-ethical) method, which increase the influence of a promoted message in social networks by the construction of Spreading Groups.

These Spreading Groups consist of a sub set of users in the network, which ignite the spread. They detonate the process of information cascades, by passing the message between members in the group at the initial stages, such that the 'natural threshold' of the viral process is exceeded.

The work presents some of the theoretical advantage that can be obtained by such Spreading Groups, and follows by an empirical work based on Twitter real data, suggesting a possible existence of such groups in Twitter financial stocks information.

The spread of information through spreading groups does not require harnessing the influencers' and opinion leaders. It can be performed by humans, AI bots, or human-AI bots combinations. The advantage and potential of humans – bots spreading groups is clear.

**REFERENCES**

**Anger Isabel and Kittl Christian,** Measuring influence on Twitter, *Proceedings of the 11th International Conference on Knowledge Management and Knowledge Technologies*. - 2011. - p. 31.
**Antweiler W. and Frank M. Z.** Is all that talk just noise? The information content of internet stock message boards, T*he Journal of Finance*. - 2004. - 3 : Vol. 59. - pp. 1259-1294.
**Barabasi Albert-Laszlo and Albert Reka,** Emergence of scaling in random networks,  *Science*. - 1999. - 5439 : Vol. 286. - pp. 509-512.
**Bennett W. L. and Segelberg A.,**  Digital media and the personalization of collective action: social technology and the organization of protests ageins the global economic crisis., *Information, Communication & Societ*y. - 2011. - 6 : Vol. 14. - pp. 770-799.
**Bollobas Bela** *Random Graphs* , Cambridge : Cambridge University Press, 2001.
**Cao H. H., Coval J. D. and Hirshleifer D.,** Sidelined investors, trading-generated news, and security returns, *Reviews of Financial Studies*. - 2002. - 2 : Vol. 15. - pp. 615-648.
**Commission U.S. Securities and Exchange,**  SEC Charges 44 Stock Promoters in First Internet Securities Fraud Sweep [Online] // U.S. Securities and Exchange Commission. - US Securities and Exchange Commission, 10 28, 1998. - http://www.sec.gov/news/headlines/netfraud.htm.
**Davis Clayton A [et al.],** BotOrNot: A System to Evaluate Social Bots, arXiv preprint arXiv:1602.00975. - 2016.
**DeMarzo Peter. M., Vayanos Dimitry and Zwiebel Jeffrey,** Persuation bias, social influence, and unidimensional opinions , *The Quarterly Journal of Economics*. - 2003. - 3 : Vol. 118. - pp. 909-968.
**Freitas Carlos [et al.],**  Reverse engineering socialbot infiltration strategies in twitter, *Proceedings of the 2015 IEEE/ACM International Conference on Advances in Social Networks Analysis and Mining* 2015. - 2015. - pp. 25-32.
**Harris M. and Raviv A.** Differences of opinion make a horse race, *Review of Financial Studies.* - 1993. - 3 : Vol. 6. - pp. 473-506.
**Hirshleifer David and Teoh Siew Hong,**  Limited attention, information disclosure, and financial reporting, *Journal of accounting and economics*. - [s.l.] : Elsevier, 2003. - 1 : Vol. 36. - pp. 337-386.
**Howard Ph. N. [et al.],**  Opening closed regimes: what is the role of social media during the Arab Spring?,  *Social Science Research Network*. - 2011. - February 4, 2016. - http://ssrn.com/abstract=2595096.
**Kempe D., Kleinberg J. and Tardos E.** , Maximizing the spread of influence through a social network , *Proc. 9th ACM SIGKDD Int. Conf. Knowledge Discovery and Data Mining.* - 2003. - pp. 137-146.
**Leskovec Jure , Backstrom Lars and Kleinberg Jon,**  Meme-tracking and the dynamics of the news cycle, *Proc. 15th ACM SIGKDD Int. Conf. Knowledge Discovery and Data Mining.* - 2009. - pp. 497-506.
**Newman Mark** *Networks*. An Introduction., Oxford : The Oxford University Press, 2010.
**Subrahmanian VS [et al.]** The DARPA Twitter Bot Challenge, arXiv preprint arXiv:1601.05140. - 2016.
**Weng Lillian [et al.]** Competition among memes in a world with limited attention, Scientific reports. - [s.l.] : Nature Publishing Group, 2012. - Vol. 2.
**Wilensky Uri** NetLogo [Online] // NetLogo. - 1999. - February 6, 2016. - https://ccl.northwestern.edu/netlogo.

**ONLINE APPENDIX TO:**

**SPREADING GROUPS IN TWITTER: THE FLOWS OF RUMORS ABOUT THE STOCK PRICES**

1. The questionnaires used to measure the importance of the tweets in the experimental part can be found in:
   http://goo.gl/forms/roiUtT4LpE
   http://goo.gl/forms/kJwIMmr3PE
   http://goo.gl/forms/bKcfl7wELx
   http://goo.gl/forms/zrzMrHBriy

2. The results from the MT experiment can be found [here](.).

https://drive.google.com/file/d/0Bz4ii_5xikATSEx0NW82RTA4MGM/view?usp=sharing

3. The Netlogo code for the simulations can be found [here](.).

https://drive.google.com/file/d/0Bz4ii_5xikATcEE3eHI3UWY5b0E/view?usp=sharing